\documentclass[prd,nofootinbib,preprint,superscriptaddress]{revtex4}

\usepackage[english]{babel}

\usepackage{tikz,xcolor,hyperref}
\usepackage{amsmath, amssymb, amsthm, graphicx, epsfig, fancyhdr,epsfig, slashed}

\usepackage{natbib}
\usepackage{float}

\definecolor{lime}{HTML}{A6CE39}
\DeclareRobustCommand{\orcidicon}{
	\begin{tikzpicture}
	\draw[lime, fill=lime] (0,0) 
	circle [radius=0.2] 
	node[white] {{\fontfamily{qag}\selectfont \tiny ID}};
	\draw[white, fill=white] (-0.0625,0.095) 
	circle [radius=0.007];
	\end{tikzpicture}
	\hspace{-2mm}
}

\foreach \x in {A, ..., Z}{\expandafter\xdef\csname orcid\x\endcsname{\noexpand\href{https://orcid.org/\csname orcidauthor\x\endcsname}
			{\noexpand\orcidicon}}
}

\newcommand{\be}{\begin{equation}}
\newcommand{\ee}{\end{equation}}
\newcommand{\bea}{\begin{eqnarray}}
\newcommand{\eea}{\end{eqnarray}}

\newcommand{\mt}{\mathtt{}}

\begin{document}


\title{Non-perturbative Lee-Wick gauge theory: \\ \it{Towards Confinement \& RGE with strong couplings}}

\author{Marco Frasca\orcidA{}}
\email{marcofrasca@mclink.it}
\affiliation{Rome, Italy}

\author{Anish Ghoshal\orcidB{}}
\email{anish.krrish@gmail.com}
\affiliation{Institute of Theoretical Physics, Faculty of Physics, University of Warsaw, ul. Pasteura 5, 02-093 Warsaw, Poland}

\author{Alexey S. Koshelev\orcidC{}}
\email{ak@inpcs.net}
\affiliation{
Departamento de F\'isica, Centro de Matem\'atica e Aplica\c{c}\~oes (CMA-UBI),\\
Universidade da Beira Interior, 6200 Covilh\~a, Portugal
}

\begin{abstract}

We consider a non-Abelian 
%
Lee--Wick gauge theory and 
discuss
Becchi-Rouet-Stora-Tyutin (BRST) invariance. It contains fourth-order derivative as extensions of the kinetic term, leading to massive ghosts in the theory upon quantization. 
We particularly 
provide essential clues towards
confinement conditions in strongly-coupled regimes, using the Kugo-Ojima approach, and
obtain the $\beta-$functions in the non-perturbative regimes. This is achieved using a set of exact solutions of the corresponding local theory in terms of Jacobi elliptical functions. We obtain a 
similar $\beta-$function just as for the 
ordinary
%
Yang-Mills theory but the main differences are that now, the cut-off arises naturally from the Lee-Wick heavy mass scales (M).
%
We show that the fate of the ghosts are fixed in these regimes: they are no more the propagating degrees of freedom in the infrared (IR)-limit. 
As it also happens for the 
ordinary
case, confinement is due to the non-Abelian nature of the theory.
In the limit $M \rightarrow \infty $, one recovers the standard results for the local non-Abelian Yang-Mills theory.

\end{abstract}

\maketitle

\section{Introduction}

The search for a consistent
theory of quantum gravity has proven to be one of the main challenges of theoretical physics. Some extended formulations have been provided through string theory and loop quantum gravity without yielding a general accepted scenario. Most questions are open yet. Different avenues have been explored trying to make the Einstein theory of gravity renormalizable in the UV. A notable attempt has been the Stelle's theory where higher order derivative of the metric are added to the well-known Einstein-Hilbert action \cite{Stelle:1976gc,Stelle:1977ry,Lu:2015psa}. This theory proved to be renormalizable but has severe problems due to the presence of ghosts that break unitarity. Since then, several attempts have been made to overcome such a problem (e.g. see \cite{Calcagni:2017sdq,Buchbinder:2021wzv} and refs. therein). Anyway, one can recognize here an important problem of quantum field theory and the main model having similar characteristic as Stelle's theory is the so called Lee-Wick model \cite{LeeWick:1969} that displays identical good and bad behavior.

Stydies of the Standard Model have permitted to get formulations that can provide a deep understanding of this kind of problems as in
%
Refs. \cite{Grinstein:2007mp,Grinstein:2007iz,Wise:2009mi} where higher-order derivatives augment the SM via TeV-scale Lee--Wick partner particles with negative-sign kinetic and mass terms. This evidently leads to a cancellation of quadratic divergences in scalar self-energies, with the predictions of the  
precise particle spectrum of Lee--Wick resonances at the laboratory experiments
 to determine.\footnote{It is worthwhile to point out that also the Abelian Lee-Wick theory has a consistent behavior in UV, granting some quantities and Feynman diagrams to be finite\cite{Accioly:2011zz,Turcati:2014eaa,Barone:2015sfa,Ji:2019phv}.} 
The theory proves to be unitary, \cite{Cutkosky:1969fq} (see also 
\cite{Anselmi:2017yux,Anselmi:2017lia}
), is causal at the macroscopic level \cite{Grinstein:2008bg}, and has subsequently been generalized to include additional partner particles \cite{Carone:2008bs,Carone:2008iw}. It has received much attention in the BSM particle physics phenomenology literature as well \cite{LWpheno}. So far, no Lee--Wick partner particles have been found hitherto, thereby putting a lower bound from $\sqrt{s} \sim \mathcal{O}(10)\,\text{TeV}$ \cite{Zyla:2020zbs}, on the Lee-Wick partner masses and pushing them to higher energies.

Looking for other possible avenues to a formulation of quantum gravity,
the last decades have also seen the development of general non-local field theories, in particular ghost-free ones \cite{Efimov:1967,Krasnikov:1987,Kuzmin:1989,Tomboulis:1997gg,Modesto:2011kw,Biswas:2011ar,Wataghin:1934ann,Pauli:1949zm,Pais:1950za}, and in the context to p-adic string theory and non-commutative geometry \cite{Witten:1986,Frampton:1988,Tseytlin:1995uq,Siegel:2003vt}.\footnote{For cosmology of these theories, see Ref. \cite{Biswas:2005qr}, nonlocality in string theory see \cite{Calcagni:2013eua,Calcagni:2014vxa}, regularization of the gravitational field via nonlocality see \cite{Edholm:2016hbt,Boos:2018bxf,Giacchini:2018wlf} on one hand, and on the other hand, the role of the Wick rotation vis-\`a-vis unitarity and causality \cite{Carone:2016eyp,Tomboulis:2015esa,Shapiro:2015uxa,Briscese:2018oyx,Briscese:2021mob,Koshelev:2021orf}.} Recently, in  this context, higher-derivative approaches to a UV-completion of QFT have become popular\cite{Moffat:1990jj, Evens:1990wf, Tomboulis:1997gg, Moffat:2011an, Tomboulis:2015gfa,Kleppe:1991rv}, and particularly, the infinite higher-derivative approach that was motivated starting from string field theory \cite{sft1,sft2,sft3,padic1,padic2,padic3,Frampton-padic,marc,Tseytlin:1995uq,Siegel:2003vt,Calcagni:2013eua,Modesto:2011kw,Modesto:2012ga,Modesto:2015foa,Modesto:2017hzl} where attempts were made to address the divergence problem by generalizing the kinetic energy operators of the Standard Model (SM) to an infinite series of higher order derivatives, suppressed by the scale of non-locality ($M$) at which the higher order derivatives come into the picture \cite{Krasnikov:1987yj,Biswas:2014yia}, also to readily cure the vacuum instability problem in the SM \cite{Ghoshal:2017egr}. Such a theory is ghost-free 
\cite{Buoninfante:2018mre}, predicts conformal invariance in the ultra-violet (UV), trans-planckian scale transmutation and dark matter phenomenology \cite{Ghoshal:2018gpq,Buoninfante:2018gce} and is free of Landau poles making it a candidate theory for UV-completion of 4D QFT, valid and perturbative up to infinite energy scales \cite{Ghoshal:2017egr,Ghoshal:2018gpq,Ghoshal:2020lfd}. Strong coupling regimes of the theory were studied in Refs. \cite{Frasca:2020jbe,Frasca:2020ojd,Frasca:2021iip}, where it was shown that the mass gap obtained gets diluted in the UV due to non-local effects restoring conformal invariance in the UV along with studies of false vacuum \cite{Ghoshal:2022mnj} and Higgs dark energy \cite{Frasca:2022vvp}. Interestingly, it was shown that Lee--Wick theory (Higgs, abelian and non-Abelian gauge theories) with $N$ propagator poles and having $(N-1)$ Lee--Wick partners can be understood as flowing to infinite-derivatives in the $N \rightarrow \infty $ asymptotic limit \cite{Boos:2021chb,Boos:2021jih,Boos:2021lsj}.


Renormalization group equations (RGEs) helps us to understand the relevance of UV fixed points for quantum field theories \cite{Wilson:1971bg,Wilson:1971dh}, for example, in quantum chromodynamics. Here, the property of the asymptotic freedom manifests the reliability of the theory by the use of the standard perturbation theory \cite{Gross:1973id,Politzer:1973fx,Callaway:1988ya}. Now, when the fixed point corresponds to an interacting theory, we dub this as asymptotic safety \cite{Weinberg:1980gg}.  It may be desirable for UV-completion theories, that are neither asymptotically free nor renormalizable, to have such a UV fixed point behaviour \cite{Litim:2011cp}. This idea was recently developed for quantum gravity \cite{Litim:2011cp,Litim:2006dx,Niedermaier:2006ns,Niedermaier:2006wt,
Percacci:2007sz,Litim:2008tt,Reuter:2012id}. Applications to the Standard Model have also shown the possibility of such a UV fixed point \cite{Litim:2014uca,Giudice:2014tma,Pelaggi:2017abg,Mann:2017wzh,Antipin:2018zdg,Molinaro:2018kjz}. The authors studied the RGEs of non-local infinite-derivative theories in Ref. \cite{Ghoshal:2017egr,Ghoshal:2020lfd,Frasca:2021iip}, but however the story remains unclear for Lee--Wick theories, particularly in the non-perturbative regimes which we investigate in this paper.

Besides the RGE approach in QFT, deeper understanding of the confinement of quarks in the Standard Model (QCD sector) (see \cite{Kogut:2004su} and references there-in) was proposed by  Kugo and Ojima that firstly proposed a confinement condition based on the BRST invariance and charge annihilation and then, later on, extended to color confinement by other authors \cite{Kugo:1977zq,Kugo:1979gm,Nishijima:1993fq,Nishijima:1995ie,Chaichian:2000sf,Chaichian:2005vt,Nishijima:2007ry,Seiberg:1994aj,Seiberg:1994rs,Chaichian:1999is}.\footnote{Studies by Gribov \cite{Gribov:1977wm} and Zwanziger \cite{Zwanziger:1989mf} suggested confinement in QCD with the gluon propagator running to zero as momenta go to zero and an enhanced ghost propagator running to infinity more rapidly than the free case in the same limit of momenta. This scenario was not confirmed by studies of the gluon and ghost propagators on the lattice \cite{Bogolubsky:2007ud,Cucchieri:2007md,Oliveira:2007px}. Indeed, the existence of a mass gap was proven unequivocally in lattice computations for the spectrum of Yang-Mills theory without fermions \cite{Lucini:2004my,Chen:2005mg}.
These results have found theoretical basis in Refs. \cite{Cornwall:1981zr,Cornwall:2010bk,Dudal:2008sp,Frasca:2007uz,Frasca:2009yp,Frasca:2015yva} in terms of a closed form formula for the gluon propagator (see Ref. \cite{Deur:2016tte} for a review).}  Confinement, in its simplest form, can be understood as the combined effect of a potential obtained from the Wilson loop of a Yang--Mills theory without fermions and the running coupling yielding a linearly increasing potential, in agreement with lattice data \cite{Deur:2016bwq}. 

The main motivation for this work is to present a completely new approach that can help the formulation of a consistent theory of quantum gravity in the UV limit. Indeed,
in this work, we will apply the conditions of confinement derived in \cite{Nishijima:1993fq,Nishijima:1995ie,Chaichian:2000sf,Chaichian:2005vt,Nishijima:2007ry} ( via reducing it to the case of the Kugo--Ojima criterion \cite{Kugo:1979gm} ) to non-local non-Abelian Lee--Wick gauge theories without including fermions, following similar approach as studied in the case of ghost-free infinite-derivative non-local Yang-Mills \cite{Frasca:2021iip}. This uses the form of exact solutions of the background using elliptical functions and the technique has been successfully applied to calculate non-perturbative local theory phenomenology: Hadronic contribution of muon (g-2)$_{\mu}$ \cite{Frasca:2021yuu} and tunneling of the false vacuum \cite{Frasca:2022kfy,Calcagni:2022tls} 
We will get several clues pointing out to a confined non-Abelian Lee-Wick gauge theory in 4-D and a mass gap is obtained.
%
We show that the Lee--Wick higher-derivative operators defined in the UV yields finite contributions also in the IR-limit and provide 
clues towards
a proof of confinement, granted by the absence of the Landau pole. 
We emphasize that, also in the 
Lee-Wick
case, it is the non-Abelian nature of the theory that supports confinement and not the 
higher-order derivatives
due to the Lee-Wick model. 
Lee-wick quantum field theory, which serves as toy model scenario for Stelle's theory of quantum gravity, also carries the pathological ghosts and in this paper we show that in the strongly-coupled limit the theory acquires a mass gap and possibly the Lee-Wick ghosts gets confined and do not appear as propagating degrees of freedom in the theory.\footnote{
Generic complex ghosts were also shown to be confined recently in Ref. \cite{Frasca:2022gdz}}

These results can be framed in the recent reformulation of the Lee-Wick theory in perturbative regime, due to Anselmi and Piva, where unphysical degrees of freedom are seen to not propagate \cite{Anselmi:2017yux,Anselmi:2017lia}. We show that the same conclusions can be reached in the non-perturbative case when the IR limit is considered.

The paper is organized as follows: in section II, we review Lee-Wick gauge theory
In section III, we introduce BRST transformation in Lee--Wick gauge theory, and prove that the theory is BRST-invariant. Following this, in section IV, we 
discuss
confinement of Lee--Wick gauge field in the strongly-coupled regimes and derive the exact RGE in this context. Next in section V, 
we derive the beta function
and we end by discussing the salient conclusions drawn from our study in the last section.

\medskip

\section{Lee-Wick Gauge Theories}
\label{Non_Abelian}

We know the action for the SU(N) pure Yang-Mills theory, in the 
ordinary
%
case, takes the form 
\be
\label{eq:YMloc}
L_{loc}=-\frac{1}{4}  F^{a\mu\nu} F_{a\mu\nu}.
\ee
where the repeated indexes imply summation both for space-time and group indexes and the field strength tensor $F^{a\mu\nu}$ is given by
\be
F_{\mu\nu}^a=\partial_{[\mu}A_{\nu]}^a -gf^{abc}A^b_{\mu}A^c_{\nu}     \ ,
\ee
with the group structure constants is denoted by $f^{abc}$ and the dimensionless gauge coupling is $g$. 
We extend the theory to the 
higher-derivative 
%
case by following the approach given in Refs.~\cite{Ghoshal:2017egr,Ghoshal:2020lfd}. When involving higher-derivative extensions of a field theory, the free part of the theory can be written as \cite{Krasnikov:1987yj,Biswas:2014yia,Ghoshal:2017egr,Ghoshal:2018gpq,Ghoshal:2020lfd}:
\begin{equation}
    L_{nloc} = -\frac{1}{4}F^a_{\mu\nu}U(D^2)F^{a\mu\nu}.
\end{equation}
where $U(D^2)$ is some function of $D^2$, and 
\be
\label{eq:D}
D_\mu^{ab}=\partial_\mu \delta^{ab}-igA^c_{\mu}(T^{c})^{ab}
\ee
is the covariant derivative in the adjoint representation. We have introduced a mass scale $M$ for the scale of new physics. Here in this case, it is where the higher-order derivatives come into play. 



Particularly, in the Lee--Wick model, the proper choice of $U(D^2)$ is granted by
\be
U(D^2)=I+\frac{D^2}{M^2}.
\ee
The total $\mathcal{\rm Lagrangian}$ is then given by
\be
\label{eq:fullL}
L=L_{nloc}+L_{gf}+L_{ghost}.
\ee
where
\be
L_{gf}=\frac{\xi}{2}B^2+B^a\partial^\mu A_\mu^a,
\ee
with $B^a$ an auxiliary field, and
\be
L_{ghost}={\bar c}^a(-\partial^\mu U_1(\Box)D^{ac}_\mu)c^c.
\ee
For Lee--Wick gauge theory extensions, let us choose the function, involving higher-order derivatives,
\be
U_1 (\Box)=1-\frac{\Box}{2M^2}.
\ee
Please note that both the functions $U$ and $U_1$ are chosen in such a way that, in the limit $M\rightarrow\infty$ the 
standard
theory is properly recovered \footnote{Throughout the paper we will work in the Euclidean metric as done in other work in the literature (see references for the Lee-Wick theory in the introduction).}.

%
In this theory, the unperturbed sector of the Lagrangian displays a ghost field that could freely propagate. This could give rise to anomalies when the theory is properly quantized. In the Appendix A, we show that, already at a classical level, the ghost field could be proven to be ineffective in view of a non-perturbative approach. This is somewhat preparatory to our confinement proof.

\medskip

\section{Confinement \& BRST Invariance in Yang-Mills}
\label{BRST}


In the following sections, we present the confinement condition as devised in \cite{Nishijima:1993fq,Nishijima:1995ie,Chaichian:2000sf,Chaichian:2005vt,Nishijima:2007ry}, reducing it to the Kugo-Ojima criterion \cite{Kugo:1979gm}, projecting all the discussion on the exact solutions obtained in \cite{Frasca:2015yva,Frasca:2020jbe,Frasca:2020ojd}.
%
This will permit us to draw some conclusions on a viable proof of confinement for gauge theory, both in 
the ordinary and Lee-Wick cases.
%

\subsection{
Yang-Mills
Theory}


The formalism we present here for the local case is the same as used in \cite{Chaichian:2018cyv}. The Yang-Mills Lagrangian can be subdivided in three parts as follows
\begin{equation}
\label{lagrangian}
L=L_{{loc}}+L_{{gf}}+L_{{ghost}}.
\end{equation}
%
We have $L_{loc}$ for the classical gauge-invariant part, ${L}_{{gf}}$ for the gauge-fixing terms and ${L}_{{ghost}}$ for the Faddeev--Popov (FP) ghost term 
%
and
\begin{eqnarray}
\label{eq:L2} 
%
L_{{gf}}&=&\partial_\mu B^a A^{a\mu}+\frac{1}{2}\xi B^a
B^a\,,\cr
L_{{ghost}}&=&i\partial_\mu \bar c^a D^{ab\mu}
c^b\,,\label{lagr_terms} 
\end{eqnarray}
%
where $\xi$ denotes the gauge parameter and $D_\mu$ is the covariant derivative already defined in eq.~(\ref{eq:D}).
%
%

BRST transformations for a generic field $\chi$ can be expressed by BRST charges $Q_B$ and $\bar Q_B$ given by \cite{Kugo:1977zq}
\begin{eqnarray} 
\delta\,\chi=i[Q_B,\chi]_\mp,\ \ \ \bar\delta\,\chi=i[\bar
Q_B,\chi]_\mp\,,\\\label{brs_chi}
Q_B^2={\bar Q}_B^2=Q_B\bar Q_B+\bar Q_BQ_B=0\,.\label{brs_charge}
\end{eqnarray}
We will take the $-(+)$ sign in (\ref{brs_chi}) when $\chi$ is even (odd) in the ghost fields $c$ and $\bar c$. We recognize them as anti-commuting scalar fields.


The BRST transformations can be generally defined in the following way
%
\bea
\delta A_\mu^a&=&D_\mu^{ab} c^b\,,\cr
%
\bar\delta A_\mu^a&=&D_\mu^{ab} \bar c^b\,,
\label{brs_transf1} 
\eea
By imposing for the auxiliary fields $B$, $c$ and $\bar c$ 
\begin{equation} 
\delta{\cal L}=\bar\delta{\cal L}=0\,, 
\end{equation}
one gets
\begin{eqnarray} 
\delta\,B^a=0\,,\ \ \ \delta\,\bar c^a=i B^a\,,\ \ \
\delta\,c^a=-\frac{1}{2}gf^{abc} \,(c^b c^c)\,,\cr
\bar\delta\,\bar B^a=0\,,\ \ \ \bar\delta\,c^a=i \bar B^a\,,\ \ \
\bar\delta\,\bar c^a=-\frac{1}{2}gf^{abc}\,(\bar c^a\bar c^c)\,, 
\end{eqnarray}
with $\bar B^a$ defined by the following equation
\begin{equation} 
B^a+\bar B^a-igf^{abc}(c^b\bar c^c)=0\,.
\end{equation}
%

By a direct application of the Noether's theorem, one has a conserved current given by
\begin{equation}
j_{\mu}=\sum_{\{\Phi\}}\frac{\partial \ L}{\partial (\partial_\mu \Phi)}\delta\Phi
=B^a(D_\mu c)^a -\partial_\mu B^a c^a+i\frac12{\rm g}f^{abc}\partial_\mu \bar c^a c^bc^c,
\end{equation}
with $\{\Phi\}$ being the set of all fields present in the Lagrangian. Therefore, the corresponding charge $Q_B$ is given by
\begin{equation}
Q_B=\int d^3x \left(B^a(D_0 c)^a -\dot B^a c^a+i\frac12{\rm g}f^{abc}\dot{\bar c}^a c^bc^c\right).
\end{equation}

Therefore, we have
\begin{equation}
\delta(L_{{gf}}+L_{{FP}})=\delta(-i\partial_\mu\bar c^a
A_\mu^a-\frac{i}{2}\xi\,\bar c^a \cdot B^a),
\end{equation}
confirming that
\begin{equation} 
\delta{L}_{f}=0\,. 
\end{equation}
%

Given this Lagrangian, the equations of motion are given by
\begin{equation}
   D^{\mu ab}F^b_{\mu\nu}+j^a_\nu=i\delta\bar\delta A^a_\nu.
\end{equation}
The contributions of the auxiliary fields are on the right-hand side.
These represent massless particles at tree level.
It is also easy to see that the $B$ field is not propagating.
The consequences of this is that such fields will not give any contribution to the physical spectrum of the theory.
In order to evaluate such a contribution, we have to compute
\begin{equation}
\label{eq:corr}
\begin{array}{lll}
\langle i \delta \bar{\delta} A_{\mu} ^{a} (x), A_{\nu} ^{b} (y)
\rangle.
\end{array}
\end{equation} 

By the Kugo-Ojima technique, one has
\begin{equation}
   \delta \bar{\delta} A_{\mu}^{a}=-\{Q_B,\{\bar{Q}_B,A_{\mu} ^{a}\}\}.
\end{equation}
Then, because of $\langle 0|Q_B=Q_B|0\rangle=\bar{Q}_B|0\rangle=\langle 0|\bar{Q}_B=0$, one has
\begin{equation}
   \langle i \delta \bar{\delta} A_{\mu}^{a} (x),A_{\nu}^{b} (y) \rangle=\langle i  \bar{\delta} A_{\mu}^{a} (x),\delta A_{\nu}^{b} (y) \rangle=
	i\langle D_\mu\bar{c}^{a} (x),D_\nu c^{b} (y) \rangle.
\end{equation}
Then, our strategy is to evaluate the above expression in momentum space assuming it in the following most general form
%
%
\begin{equation}
\label{eq:KOc}
\int d^dxe^{ipx}\langle D_\mu\bar{c}^{a} (x),D_\nu c^{b} (y) \rangle=\delta^{ab}
\left(\delta_{\mu\nu} - \frac{p_{\mu} p_{\nu}}{p^2-i\epsilon}\right) u(p^2)-\delta^{ab}\frac{p_{\mu} p_{\nu}}{p^2-i\epsilon},
\end{equation}
This means that the the no-pole condition
%
takes the form given
here 
\cite{Kugo:1979gm}
\begin{equation} 
\label{eq:KO}
1+u(p^2=0)=0,
\end{equation}
which is the Kugo--Ojima condition for confinement granting that no massless pole appears in the spectrum of the theory. Indeed, this condition removes the massless term from Eqn.(\ref{eq:KOc}).
We just notice that the value of our function $u(p^2)$ in 0 coincides with the parameter $u$ in Ref.~\cite{Kugo:1979gm}.



The form of the $u(p^2)$ function has been given explicitly in \cite{Chaichian:2018cyv} and, from Appendix B in eq.(\ref{eq:uu}), it is seen to be
\be
\label{eq:uu0}
u(p^2)=\frac{(N^2-1)^2}{2N}g^2\delta^{ab}
\left(\delta_{\mu\nu} - \frac{p_{\mu} p_{\nu}}{p^{2}}\right)
\int\frac{d^4p'}{(2\pi)^4}K_2(p-p')G_2(p').
\ee
%
The $\beta$-function was also obtained and given by eq.(\ref{eq:betaYM}) in Appendix B. Below, we will extend this method to the Lee-Wick theory.

\medskip

\subsection{Lee--Wick Gauge Theory}

We present a brief discussion of BRST symmetry in Lee-Wick gauge theory in order to make the paper self-contained and better understandable. Anyway, some discussion about can be found in \cite{Ghoshal:2020lfd}.
We would like to point out that, for Lee-Wick gauge theories, a set of Ward identities holds\cite{Grinstein:2007iz}. The use of this technique in our case entails the need for Faddeev-Slavnov identities that should be proven to hold also in the Lee-Wick non-Abelian case.


In order to make clear the extent to which Kugo-Ojima formalism applies to Lee-Wick theory, we just point out that in Ref.~\cite{Ghoshal:2020lfd} BRST symmetry has been extended to non-local infinite derivative theories. The real modification in the BRST transformation has been shown minimal, as we will see below, and so, there is no problem to extend the Kugo-Ojima formalism through a proper definition of charges. In this work, we derive the Lee-Wick theory as a series expansion of the form factor of the infinite derivative theory and, as such, the same argument should apply in this case for the Kugo-Ojima formalism. If such a formalism appears to be applicable in such a case, the immediate consequence is that all the confinement arguments implied by it apply as well. This should not be seen as a full formal proof but it is our understanding of why all this approach appears to work in the Lee-Wick case and the infinite derivative theories as well. Indeed, for the latter case, confinement was discussed in \cite{Frasca:2021iip}.

The complete BRST-invariant infinite-derivative gauge theory in the quantized action has the form \cite{Ghoshal:2020lfd}: 
\be \label{F1}
L = L_{nloc} + \frac{\xi}{2} (B^a) ^2 + B^a \partial ^{\mu} A_{\mu}^a + \bar{c} ^a (-\partial ^{\mu} U_1(\Box) D_{\mu}^{ac})c^c,
\ee
where $\xi$ is the gauge fixing parameter, $B$ is the auxiliary field, and $c$ and $\bar{c}$ are the ghost and anti-ghost fields, respectively.
The BRST transformations for non-Abelian gauge theories express a residual symmetry
of the effective action which remains after the original gauge invariance has been
broken by the addition of the gauge-fixing and ghost action terms. 
Our BRST transformations are modified in the following way:

\begin{eqnarray}
\label{eq:brst1} 
\delta A_\mu^a&=&D_\mu^{ab} c^b\,,\cr
\bar\delta A_\mu^a&=&D_\mu^{ab} \bar c^b\,,\label{brs_transf} 
\end{eqnarray}
and
\begin{eqnarray} 
\label{eq:brst2}
\delta\,B^a=0\,,\ \ \ \delta\,\bar c^a=i U_1^{-1}(\Box)B^a\,,\ \ \
\delta\,c^a=-\frac{1}{2}gf^{abc} \,(c^b c^c)\,,\cr
\bar\delta\,\bar B^a=0\,,\ \ \ \bar\delta\,c^a=iU_1^{-1}(\Box)\bar B^a\,,\ \ \
\bar\delta\,\bar c^a=-\frac{1}{2}gf^{abc}\,(\bar c^b \bar c^c)\,.
\end{eqnarray}
We show the BRST-invariance of 
$S _ {\mt{inv} }$ by noting that
the BRST transformation of the gauge field is just a gauge transformation of $A^a_{\mu}$ generated
by $c_{a}$ or ${\bar c}_{a}$
. Therefore, any gauge-invariant functionals of $F_{\mu \nu}$, 
like the first term in Eqn.~(\ref{F1}) gives
$\delta L_f = 0.$
The second term in Eqn.~(\ref{F1}) gives $\delta(\frac{\xi}{2} (B^a) ^2) = 0$ from Eqn.~(\ref{eq:brst2}).
For the third term in Eqn.~(\ref{F1}), the transformation of A$_{\mu}^a$ cancels the transformation of $\bar{c}^a$ in the last term due to Eqs.~(\ref{eq:brst1}),  leaving us with
\be \label{F2}
\delta(D^{ac} _{\mu} c^c) = D^{ac} _{\mu} \delta c^c + g f^{a b c} \delta A ^b_{\mu} c^c,
\ee
 which is is equal to 0, using the properties of the covariant derivative and the \textit{Jacobi identity} (see Ref.~\cite{Peskin:1995ev}).
The transformation of $c^{\sigma}$ is nilpotent,
\be \label{F20}
\delta(\partial_{\mu} c^{a}c^{b})=0\,,
\ee
while the transformation of $A^{a\mu}$ is also nilpotent,
\be \label{F30}
\delta((D_{b}^{\mu})^a c^{b}) = 0\,.
\ee
Hence, the action in Eqn.~(\ref{F1}) is BRST-invariant.
Noting the fact that the only part of the ghost action which varies under the BRST transformations is that of the anti-ghost ($\bar{c} _{a}$), the central idea behind our proof of BRST-invariance is that we have chosen the BRST variation of the anti-ghost ($\bar{c}_{a}$) (see Eqs.~(\ref{eq:brst2})) to cancel the variation of the gauge-fixing term.


It is not difficult to see that, in the limit of the 
mass scale
$M\rightarrow\infty$, the BRST transformations given in Eqn.(\ref{eq:brst1})-(\ref{eq:brst2}) become identical to those of the 
Yang-Mills
case. Formally, the confinement condition of Eqn.(\ref{eq:KO}) remains untouched as the effects of the non-locality, if present, are kept into the $u$ function.

\bigskip

\section{Condition of Confinement in Lee--Wick Theory}
\label{KO_conf}


In this section, we derive the confinement for the 
Lee--Wick
theory, following Ref.~\cite{Chaichian:2018cyv}. See 
the appendix B
for a brief
and more formal
review of this technique.
We start with the main assumption given in the previous section about the Kugo-Ojima formalism when applied to the Lee-Wick theory. With such a {\it proviso}, we repeat the same argument presened in Ref.~\cite{Chaichian:2018cyv} for the ordinary Yang-Mills theory with the proper modifications due to the presence of higher-derivatives as already discussed previously.
%

To summarize our strategy, we evaluate the function $u(p^2)$ and apply the Kugo-Ojima condition $u(p^2=0)=-1$ as, in our scheme, $u(p^2=0)$ is exactly their $u$ parameter whose value determines confinement that we extend below to the 
Lee--Wick
case. Finally, we are able in this way to obtain the $\beta$-function of the theory from which we can draw some conclusions about the behavior of the running coupling and confinement. It should be pointed out that we are assuming that Kugo-Ojima formalism for BRST is mathematically well-acquired and we perform some renormalization at some stage to get our result.

From the action (\ref{F1}), we derive the equations of motion,
\be
U(D^2)D^\mu F^a_{\mu\nu}+j_\nu^a=i\delta{\bar\delta}A_\nu^a.
\ee
The RHS can be evaluated as already done for the 
Yang-Mills
case, and we write down
\be
\int d^4x e^{ipx}\langle D_\mu{\bar c}^a(x),D_\nu c^b(0)\rangle=\delta_{ab}\left(\delta_{\mu\nu}-\dfrac{p_\mu p_\nu}{p^2}\right)u(p^2)-\delta_{ab}\frac{p_\mu p_\nu}{p^2}U_1(-p^2).
\ee
Indeed, this is the most general form for the given correlation function but, for the massless contribution, we have also to take into account the contribution of the 
higher-order derivatives
. The interesting part here is that all 
such
contributions enters into the definition of the function $u$.
They
arise from the two-point functions of the 
Lee--Wick
theory and we note that the fluctuations from UV can yield a significant contribution to confinement as they are summed up in the integral where they cannot be neglected. Then, the confinement condition is again
\be
1+u(p^2=0)=0.
\ee
\medskip

\subsection{Confinement in Lee--Wick gauge theory}
\label{green}

In our preceding works, we obtained the 2P-functions for infinite-derivative Yang-Mills theory \cite{Frasca:2020ojd}. 
We have for the 2P-function for the gluon field in the Landau gauge
\be
D_{\mu\nu}^{ab}(p)=\left(\eta_{\mu\nu}-\frac{p_\mu p_\nu}{p^2}\right)G_2(p),
\ee
being
\be
\label{eq:G2s}
G_2(p)=\frac{e^{\frac{1}{2}f(-p^2)}}{p^2+\Delta m^2e^{\frac{1}{2}f(-p^2)}}\frac{1}{1-\Pi(p)},
\ee
where $e^{\frac{1}{2}f(-p^2)}$ is the Fourier transform of the non-locality form factor of the theory for the infinite-derivative case as given in Ref.~\cite{Frasca:2020ojd}.
In the local limit, $M\rightarrow\infty$, Eqn.(\ref{eq:G2s}) reduces to a Yukawa form that yields a fair approximation to the exact local propagator obtained in \cite{Frasca:2015yva}. In the non-local case, one has the mass gap
\be
\label{eq:mg}
\Delta m^2=
\mu^2
 \left(18Ng^2\right)^\frac{1}{2}\frac{4\pi^2}{K^2(i)}
\frac{e^{-\pi}}{(1+e^{-\pi})^2}e^{f\left(-\frac{\pi^2}{4K^2(i)}\sqrt{\frac{Ng^2}{2}}\mu^2\right)}
 +\delta m^2,
\ee
where $K(i)=1.31102877714\ldots$ is the complete elliptical integral of the first kind\footnote{We use the notation $K(i)$ because we work with the modulus (this choice can be found in \cite{NIST}). Another notation is also possible using the square of the modulus and, in such a case, one should write $K(-1)$ having identical numerical value as given in the main text. This latter choice is adopted by some computer algebra systems like Mathematica by Wolfram Research, Inc.}.
This must be completed by the gap equation
\be
\label{eq:mgge}
\delta m^2=2Ng^2G_2(0)=2Ng^2\int\frac{d^4p}{(2\pi)^4}G_2(p).
\ee
The function $\Pi(p)$ can be neglected as also the shift $\delta m^2$ as a first approximation. Similarly, for the ghost one has
\be
K_2(p)=-\frac{1}{p^2}e^{\frac{1}{2}f(-p^2)}.
\ee
Therefore, given the following integral from eq.(\ref{eq:KOc}) and eq.(\ref{eq:uu0})
%
\begin{eqnarray}
\int d^4xe^{ipx}\langle D_\mu\bar{c}^{a} (x),D_\nu c^{b} (0) \rangle&=&-\delta^{ab}\frac{p_\mu p_\nu}{k^2}\\
&+&\frac{(N^2-1)^2}{2N}g^2\delta^{ab}
\left(\delta_{\mu\nu} - \frac{p_{\mu} p_{\nu}}{p^{2}}\right)
\int\frac{d^4p'}{(2\pi)^4}K_2(p-p')G_2(p').\nonumber
\end{eqnarray}
one gets
the confinement condition
\be
\label{eq:NL0}
u(0)=-\frac{(N^2-1)^2}{2N}g^2
\int\frac{d^4p}{(2\pi)^4}\frac{1}{p^2}
\frac{e^{f(-p^2)}}{p^2+\Delta m^2e^{\frac{1}{2}f(-p^2)}}.
\ee
So far, we have just considered a generic non-locality form factor $e^{f(-p^2)}$. To formally recover the Lee-Wick model we made the following substitution everywhere in our equations for confinement.
\be
e^{f(-p^2)}\rightarrow 1-\frac{p^2}{M^2},
\ee
that also implies
\be
e^{\frac{1}{2}f(-p^2)}\rightarrow 1-\frac{p^2}{2M^2},
\ee
We emphasize that this is a formal passage that physically entails a link between an infinite-derivative non-local model and the Lee-Wick model.
This will yield the integral
\bea
\label{eq:NL1}
u(0)&=&-\frac{(N^2-1)^2}{2N}g^2
\int\frac{d^4p}{(2\pi)^4}\frac{e^{\frac{1}{2}f(-p^2)}}{p^2}
\frac{1}{p^2e^{-\frac{1}{2}f(-p^2)}+\Delta m^2}= \nonumber \\
&&-\frac{(N^2-1)^2}{2N}g^2
\int\frac{d^4p}{(2\pi)^4}\frac{1-\frac{p^2}{2M^2}}{p^2}
\frac{1}{p^2(1+\frac{p^2}{2M^2})+\Delta m^2}.
\eea
%
We note that, even if the theory is UV-finite, not all the integrals one gets are finite themselves and the introduction of a cut-off is needed. Indeed,
in order to keep the integral UV-finite, we choose the only available energy scale in the theory i.e. $M$. Then, after performing the integral,
we will have
\be
u(0)=-\frac{(N^2-1)^2}{4\pi N}\alpha_s\left[
\frac{3}{4\sqrt{1-2z^2}}
\ln
\frac{
\left(1+\frac{1}{\sqrt{1-2z^2}}\right)
\left(1-\frac{2}{\sqrt{1-2z^2}}\right)
}
{
\left(1-\frac{1}{\sqrt{1-2z^2}}\right)
\left(1+\frac{2}{\sqrt{1-2z^2}}\right)
}
+\frac{1}{4}\ln\left(\frac{2z^2}{3+2z^2}\right)
\right],
\ee
being $\alpha_s=g^2/4\pi$. Here it is $z=\Delta m^2/M^2$. 
%
The mass gap is given by
\be
\Delta m^2\approx\mu^2\alpha_s^\frac{1}{2}\eta_0
\left(1-\eta_1\mu^2\alpha_s^\frac{1}{2}/M^2\right),
\ee
where we neglected the correction arising from the gap equation (\ref{eq:mgge}), being the numerical constants given by
\be
\eta_0=(72\pi)^\frac{1}{2}\frac{4\pi^2}{K^2(i)}\frac{e^{-\pi}}{(1+e^{-\pi})}
\ee
and
\be
\eta_1=\frac{\pi^2}{4K^2(i)}(2\pi)^\frac{1}{2}.
\ee

\section{Renormalization Group Equation}

\begin{figure}[H]
\centering
\includegraphics[height=8cm,width=10cm]{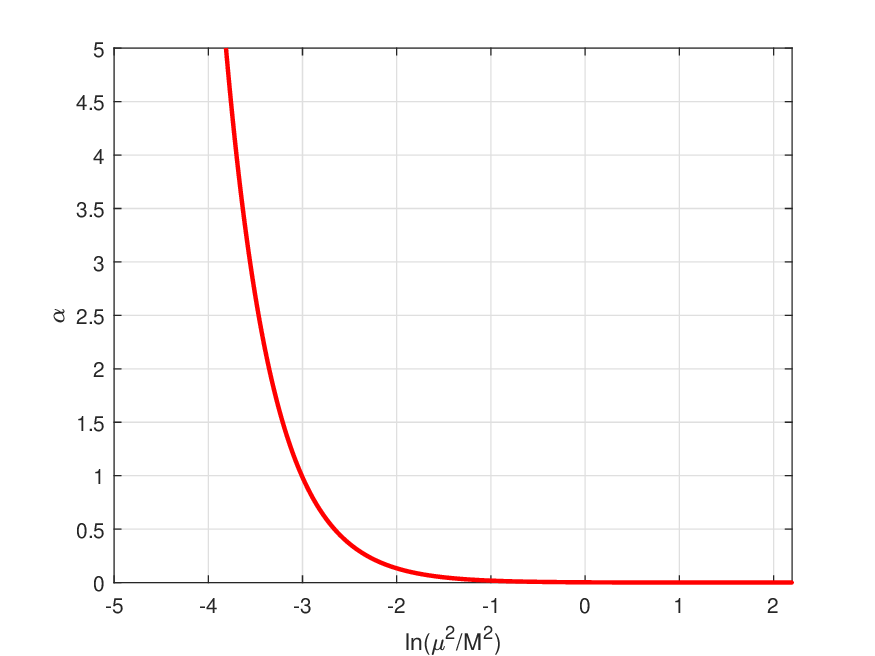}
\caption{\it Running coupling obtained from eq.(\ref{eq:alpha1}) given as a function of $\ln(\mu^2/M^2)$. Confinement arises from the coupling going to infinity in the IR limit without manifesting a Landau ghost. \label{fig1}}
\end{figure}

We can consider $\mu$ as a running mass and obtain the beta function in the limit $z\rightarrow 0$. Then, our results can be trusted only in the IR limit. We just note that
\be
z=\frac{\Delta m^2}{M^2}=\mu^2\alpha_s^\frac{1}{2}\eta_0/M^2+O(\mu^4/M^4)
\ee
so that, the leading order suffices. Therefore, observing that the function u(0) has an asymptotic leading order going like $-\ln z$, the confinement condition becomes
\be
\label{eq:alpha}
\frac{(N^2-1)^2}{4\pi N}\alpha_s\ln z=
\frac{(N^2-1)^2}{4\pi N}\alpha_s\ln\left(\mu^2\alpha_s^\frac{1}{2}\eta_0/M^2\right)
\approx -1.
\ee
This equation can be solved exactly as
\be 
\alpha=-\frac{8\pi N}{(N^2-1)^2W\left(-\frac{16\pi N\eta_0^2}{(N^2-1)^2}\frac{\mu^4}{M^4}\right)}
\ee
where $W(z)$ is the Lambert function that solves the equation $z=xe^x$. We can expand around $\mu=0$ obtaining
\be 
\label{eq:alpha1}
\alpha(\mu)=\frac{1}{2\eta_0^2}\frac{M^4}{\mu^4}-\frac{8\pi N}{(N^2-1)^2}+O(\mu^2).
\ee
The theory is seen to be IR-confining as the coupling runs to infinity for $\mu\rightarrow 0$.
In the UV-limit the theory tends to become unstable or not existing as the coupling assumes imaginary values. So, the theory appears to be well-defined just in the low-energy limit. 
Our computations are not able to recover the proper behavior of the theory in the UV-limit granting a possible comparison with 
%
Yang-Mills theory. However, we envisage that this is possible only after a proper evaluation of the mass shift (see Appendix B and Ref.~\cite{Chaichian:2018cyv}) as also shown with similar conclusions in the infinite-derivative case \cite{Frasca:2021iip}. But such an analysis is beyond the scope of the present paper. Here instead we emphasize  that the theory has a spectrum of possible confined states in the IR to dispose of the ghost following the Kugo-Ojima criterion.
In Fig.\ref{fig1}, we plot the running coupling obtained from eq.(\ref{eq:alpha1}).
This yields the beta function
\be
\label{eq:beta}
\frac{d\alpha}{dl}=\frac{\beta_0\alpha^2}{1-\frac{\beta_0}{2}\alpha}.
\ee
being $\beta_0=(N^2-1)^2/8\pi N$.
This yields an important clue that the theory is IR confining.
%
Such a conclusion, using non-perturbative arguments, confirms the analog recent results obtained in the perturbative regime \cite{Anselmi:2017yux,Anselmi:2017lia} and the theory is seen to be well-behaved also in the IR-limit. 






%

\medskip

\section{Conclusion and Discussion}
\label{conc}

We investigated strongly coupled higher-derivative Lee--Wick gauge theory
in the 4-D in context of the confinement aspects of the theory. We compared the results with that of the standard Yang-Mills theory and discussed the infrared behaviour. We presented the $\beta$-function in the strongly-coupled regime. Below, we summarize the main findings of our paper:
\begin{itemize}
    \item We showed BRST invariance of Lee--Wick Yang-Mills theory
    in the context of the Kugo-Ojima formalism
    (see eq. \ref{eq:brst1},\ref{eq:brst2},\ref{F2},\ref{F20},and \ref{F30}).
    \item We provided some essential clues towards confinement in the Lee--Wick gauge theories and showed that confinement is determined by the scale $M$ where the ghosts of the theory come into play (see eqn.(\ref{eq:alpha}) and (\ref{eq:beta})).
    \item We derived the Renormalization Group Equations in the strongly-coupled regimes of the theory and showed that the coupling runs to infinity
    in the low energy limit, without encountering the problem of Landau ghosts. 
    (see eq. (\ref{eq:alpha}))
    %
    \item The result we obtained is trustworthy both for the IR limit: 
    The ghost does not propagate in the IR-limit 
    while the UV-behavior tends to become unphysical
    and cannot be trusted.
    \item We found that the fate of Lee-Wick ghosts are sealed: they stop propagating and instead get confined and consequently do not appear as physical degrees of freedom in the strong coupling regimes.
    %
\end{itemize}

We envisage that our results 
should
shed light on more detailed understanding of confinement and $\beta$-function analysis in the framework of renormalizable quadratic gravity theories, for which Lee--Wick theory acts as a prototype. Ghosts in quadratic gravity theories
have already been speculated to be confined and do not appear as physical degrees of freedom in the mass spectrum, in analogy to quarks and gluons in QCD-like theories, in Refs. \cite{Holdom:2015kbf,Holdom:2016xfn}. In the future, we aim to extend these studies to Stelle's gravity theory and quadratic gravity in general to see if, in the UV-limit, phenomena like confinement and mass gap can appear also in this context, making possible a trustworthy theory of quantum gravity.\footnote{See Ref. \cite{conference} for a recent conference on this topic.}


\medskip 

\section{Acknowledgement}
\label{Asck}

AK is supported by FCT Portugal investigator project IF/01607/2015.
AG thank Jens Boos for careful reading of the manuscript and comments on Lee-Wick higher derivative theory.


\newpage

\section*{Appendix A: Classical behavior of the ghost field}

In a paper by Grinstein, O'Connell and Wise \cite{Grinstein:2007mp}, it was pointed out how the presence of a ghost field in the theory seems inescapable already at a classical level when an auxiliary field is introduced. This is true even if the real meaning of such a field can be really understood only at a quantum level when small perturbation theory applies. We will see that this can be proved harmless also at a classical level. These authors consider the following Lagrangian for the Lee--Wick gauge theory 
\be
\label{eq:LW}
L_{nloc}=-\frac{1}{2}F^a_{\mu\nu}F^{a\mu\nu}
+\frac{1}{M^2}D^{ab\mu}F^b_{\mu\nu}D_\rho^{ac}F^{c\rho\nu}.
\ee
This Lagrangian has identical properties as ours and also BRST ivariance so that, confinement analysis could be in principle applied here as well.

In the Lagrangian (\ref{eq:LW}), the higher-order derivative can be removed by adding an auxiliary field. We can write
\be
L_{nloc} = -\frac{1}{2}F^a_{\mu\nu}F^{a\mu\nu}-\frac{1}{2}M^2{\tilde A}_\mu^a{\tilde A}^{a\mu}+2F^{a\mu\nu}D_\mu^{ab}{\tilde A}^b_\nu.
\ee
Then, we make the change of variable $A_\mu^a\rightarrow A_\mu^a+{\tilde A}_\mu^a$ and we obtain
\bea
L_{nloc} &=& -\frac{1}{2}F^a_{\mu\nu}F^{a\mu\nu}+\frac{1}{2}(D_\mu^{ab}{\tilde A}_\nu^b-D_\nu^{ab}{\tilde A}_\mu^b)(D^{ac\mu}{\tilde A}^{c\nu}-D^{ac\nu}{\tilde A}^{c\mu}) \nonumber \\
&&+4gf^{abc}{\tilde A}_\mu^a{\tilde A}_nu^bD^{cd\mu}{\tilde A}^{d\nu}
+\frac{3}{2}g^2f^{abc}f^{cde}{\tilde A}_\mu^a{\tilde A}_nu^b{\tilde A}^{d\mu}{\tilde A}^{e\nu} \nonumber \\
&&+gf^{abc}{\tilde A}_\mu^a{\tilde A}_\nu^bF^{c\mu\nu}-\frac{1}{2}M^2{\tilde A}_\mu^a{\tilde A}^{a\mu}.
\eea
This Lagrangian yields the following classical equations of motion
\bea
&&D_\mu^{ab}F^{b\mu\nu}-gf^{abc}{\tilde A}_\mu^b(D^{cd\mu}{\tilde A}^{d\nu}-D^{cd\nu}{\tilde A}^{d\mu})
-2g^2f^{abc}f^{cde}{\tilde A}_\mu^b{\tilde A}^{d\mu}{\tilde A}^{e\nu}
-gf^{abc}D_\mu^{bd}{\tilde A}^{c\mu}{\tilde A}^{d\nu}=0 \nonumber \\
&&-D_\mu^{ab}(D^{bc\mu}{\tilde A}^{c\nu}-D^{bc\nu}{\tilde A}^{c\mu})
-2gf^{abc}{\tilde A}^b_\mu(D^{cd\mu}{\tilde A}^{d\nu}-D^{cd\nu}{\tilde A}^{d\mu})
-3g^2f^{abc}f^{cde}{\tilde A}_\mu^b{\tilde A}^{d\mu}{\tilde A}^{e\nu} \nonumber \\
&&-gf^{abc}{\tilde A}_\mu^a{\tilde A}_\nu^bF^{c\mu\nu}-M^2{\tilde A}^{a\nu}=0.
\eea
These equations can be solved by introducing two scalar fields $\phi$ and ${\tilde\phi}$, by directly applying the mapping theorem, proved in Refs.~ \cite{Frasca:2009yp,Frasca:2015yva}\footnote{This mapping theorem has been widely utilised in Refs.~\cite{Frasca:2020jbe,Frasca:2020ojd, Frasca:2021iip,Frasca:2021yuu}. }, as
\be
A_\mu^a=\eta^a_\mu\phi, \qquad {\tilde A}_\mu^a=\eta^a_\mu{\tilde\phi}
\ee
provided the constants $\eta_\mu^a$ have the properties presented in \cite{Frasca:2020ojd} and are obtained by taking for SU(2)
\begin{equation}
\eta_\mu^a=((0,1,0,0),(0,0,1,0),(0,0,0,1)),
\end{equation}
that yields
\begin{equation}
\eta_\mu^1=(0,1,0,0),\ \eta_\mu^2=(0,0,1,0),\ \eta_\mu^3=(0,0,0,1),
\end{equation}
that implies $\eta_\mu^a\eta^{a\mu}=3$. This easily generalizes to SU(N) as \begin{equation}
\label{eq:eta1}
\eta_\mu^a\eta^{a\mu}=N^2-1.
\end{equation}
Similarly, by generalizing the SU(2) case,
\begin{equation}
\eta_\mu^a\eta^{b\mu}=\delta_{ab},
\end{equation}
and
\begin{equation}
\eta_\mu^a\eta_\nu^a=\frac{1}{2}\left(g_{\mu\nu}-\delta_{\mu\nu}\right),
\end{equation}
being $g_{\mu\nu}$ the Minkowski metric and $\delta_{\mu\nu}$ the identity tensor. This mapping will yield the field equations in terms of scalar fields as:
\bea
\partial^2\phi+Ng^2\phi^3&=&2Ng^2{\tilde\phi}^3, \nonumber \\
\partial^2{\tilde\phi}+Ng^2\phi^2{\tilde\phi}+M^2{\tilde\phi}&=&3Ng^2{\tilde\phi}^3.
\eea

In order to understand the fate of the ghost field at a classical level, we make use of a set of aforementioned exact solutions that provide an understanding for Yang-Mills theory \cite{Frasca:2015yva} and were used in Ref.~\cite{Chaichian:2018cyv} to get the beta function and the spectrum of the theory in 3 and 4 dimensions in very close agreement with lattice data \cite{Frasca:2016sky,Frasca:2017slg}. This is also the form of the 1P-correlation function of the Yang-Mills theory that permits to solve the set of Dyson-Schwinger equations of the theory \cite{Frasca:2015yva}. We take for the $\phi$ field \cite{Frasca:2015yva}
\be
\label{eq:ex}
\phi(x)=\mu_0\left(2/Ng^2\right)^\frac{1}{4}\operatorname{sn}(p\cdot x+\theta,i),
\ee
where $\mu_0$ and $\theta$ are arbitrary integration constants
, with $\mu_0$ having the dimension of a mass, 
and sn is a Jacobi elliptical function.
This solution holds provided the following dispersion relation also holds
\be
p^2=\mu_0^2\sqrt{Ng^2/2}.
\ee
For the ${\tilde\phi}$ field,
we will assume the trivial solution ${\tilde\phi}=0$. 
So, even though the Green function of the ${\tilde\phi}$ field could be not trivial, it cannot propagate being 0. This conclusion should be supported by a computation of the Dyson-Schwinger set of equations as done in \cite{Frasca:2015yva} for a local non-Abelian gauge theory.
From this argument we can conclude that one of the two degrees of freedom is not relevant to the dynamics of the classical theory while the other is a legit non-linear wave excitation. We should expect a similar fate of the ghost in the quantum theory.
%

\medskip

\section*{Appendix B: Confinement in 
%
Yang-Mills theory}
\label{AppendixC}

The argument followed in the main text is just straightforwardly obtained by the approach devised in Ref. \cite{Chaichian:2018cyv}. One observes that
\begin{eqnarray}
\label{eq:uu}
\int d^4xe^{ipx}\langle D_\mu\bar{c}^{a} (x),D_\nu c^{b} (0) \rangle&=&-\delta^{ab}\frac{p_\mu p_\nu}{k^2}\\
&+&\frac{(N^2-1)^2}{2N}g^2\delta^{ab}
\left(\delta_{\mu\nu} - \frac{p_{\mu} p_{\nu}}{p^{2}}\right)
\int\frac{d^4p'}{(2\pi)^4}K_2(p-p')G_2(p').\nonumber
\end{eqnarray}
We know the form of the propagators and these are
\be
K_2(p)=-\frac{1}{p^2+i\epsilon}
\ee
for the ghost field and
\be
G_2(p)=\frac{\pi^3}{4K^3(i)}
	\sum_{n=0}^\infty\frac{e^{-(n+\frac{1}{2})\pi}}{1+e^{-(2n+1)\pi}}(2n+1)^2\frac{1}{p^2-m_n^2+i\epsilon}
\ee
for the gauge field, 
provided the mass spectrum
\be
m_n=(2n+1)\frac{\pi}{2K(i)}\left(\frac{Ng^2}{2}\right)^\frac{1}{4}\mu,
\ee
where $K(i)$ is the complete elliptical integral of the first kind and $\mu$ is one of the integration constants in the theory. 
This is a fine approximation to the full propagator as we have omitted the mass shift induced by quantum corrections. Then, the Kugo-Ojima confinement condition takes the form 
\be
\label{eq:NL}
u(0)=-\frac{(N^2-1)^2}{2N}g^2
\int\frac{d^4p}{(2\pi)^4}\frac{1}{p^2+i\epsilon}\frac{\pi^3}{4K^3(-1)}
	\sum_{n=0}^\infty\frac{e^{-(n+\frac{1}{2})\pi}}{1+e^{-(2n+1)\pi}}(2n+1)^2\frac{1}{p^2-m_n^2+i\epsilon}=-1.
\ee
This integral can be evaluated by known techniques and one obtains the $\beta$-function in closed form
\be
\label{eq:betaYM}
\beta_{YM}=-\beta_0\frac{\alpha_s^2}{1-\frac{1}{2}\beta_0\alpha_s},
\ee
with $\beta_0=(N^2-1)^2/8\pi N$. This beta function 
grants confinement with the coupling running to infinity at lowering momenta with no Landau pole. In the UV we recover the asymptotic freedom as we expected.

\bibliographystyle{unsrt}

\end{document}